# Toxicity of Carbon Nanomaterials

Arpita Adhikari[1] and Joydip Sengupta[2*]

[1]Department of Electronics and Communication Engineering, Techno Main Salt Lake, Kolkata, India

[2]Department of Electronic Science, Jogesh Chandra Chaudhuri College, Kolkata, India

[*]Corresponding author, E-mail: joydipdhruba@gmail.com

The outstanding multidisciplinary applicability of nanomaterials has paved the path for the rapid advancement of nanoscience during the last few decades. Such technological progress subsequently results in an inevitable environmental exposure of nanomaterials. Presently, nanomaterials are employed in an extensive range of commercial products. Safe and sustainable incorporation of nanomaterials in industrial products requires a profound and comprehensive understanding of their potential toxicity. Among different nanomaterials, carbon nanomaterials marked its notable superiority towards the development of state-of-art nanotechnology due to the significant contribution of each of the carbon allotropes with varied dimensionality. The zero-dimensional fullerene, one-dimensional carbon nanotube and two-dimensional graphene possess an exclusive combination of distinctive properties that are utilized in most of the nanotechnology-based products nowadays. However, potential risk factors are associated with the production as well as the use of carbon nanomaterials. Consequently, the number of studies regarding the assessment of the toxicity of these nanomaterials has increased rapidly in the past decade. This chapter will summarize the recent scientific efforts on the toxicity evaluation of different carbon nanomaterials.

## 1. Introduction

In 1959 the idea of manipulating and controlling things on a small scale was first introduced by Richard P. Feynman[1] and later in 1974, Norio Taniguchi, a professor of Tokyo University of Science first introduced the term "nanotechnology"[2]. Since then nanotechnology finds its application in all commercial sectors like medical[3,4], fire safety[5], food industry[6], transportation[7], agriculture[8], electrical transformers[9], oil and gas industry[10], renewable energies[11] etc. According to a study performed by Vance et al.[12], the use of nanotechnology-based products has increased rapidly in recent times and is expected to elevate in future in the perspective of more discoveries associated with the unique features of new nanomaterials (Fig 1).

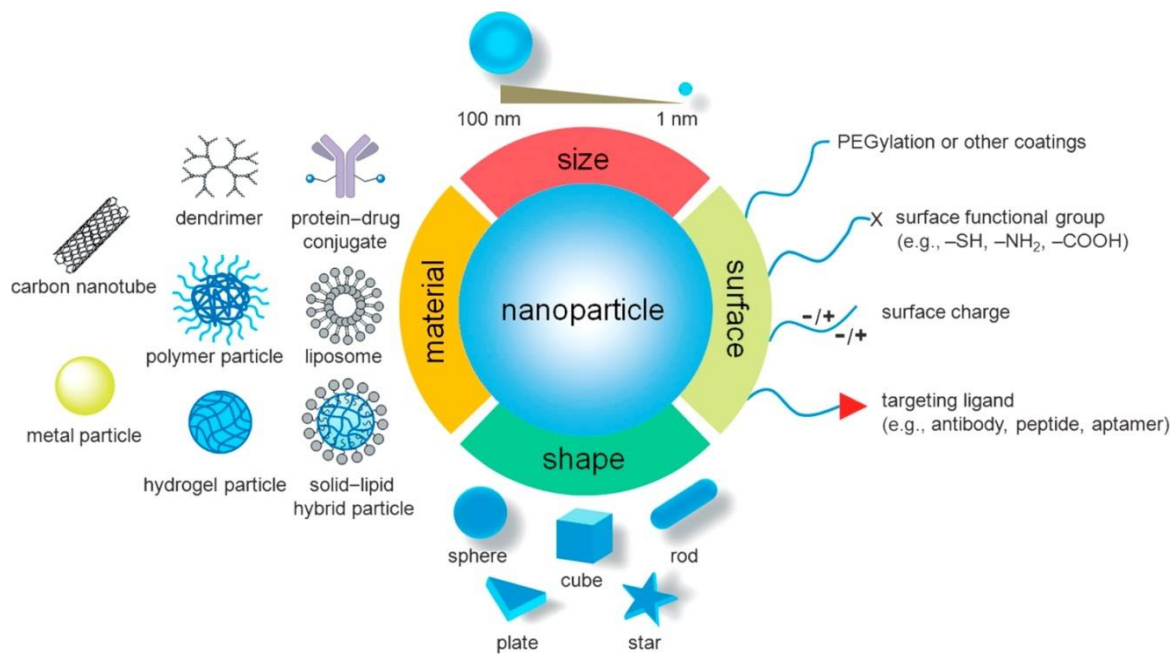

*Fig 1. Material, dimension and surface properties of nanoparticles govern their characteristics and applications[13].*

However, as the famous proverb says "*Every light has its shadow*", likewise, the excessive use of nanotechnology enhances the probability of exposure of nanomaterials towards mankind. There exist several potential routes of human exposure to nanomaterials such as ingestion, inhalation, injection and skin absorption[14] (Fig.2a). This enforces the scientists around the globe to research upon the effect of exposure of nanomaterials since way back in the 1990s[15] which consequently coined new terms like nanotoxicology[16] and nanotoxicity[17] to be evolved in 2004 and 2005, respectively. Till now the nanotoxicity of different materials [18,19,20,21] had been examined to explore the impact of their physico-chemical characteristics on the nanoparticle-biological system (Fig.2b).

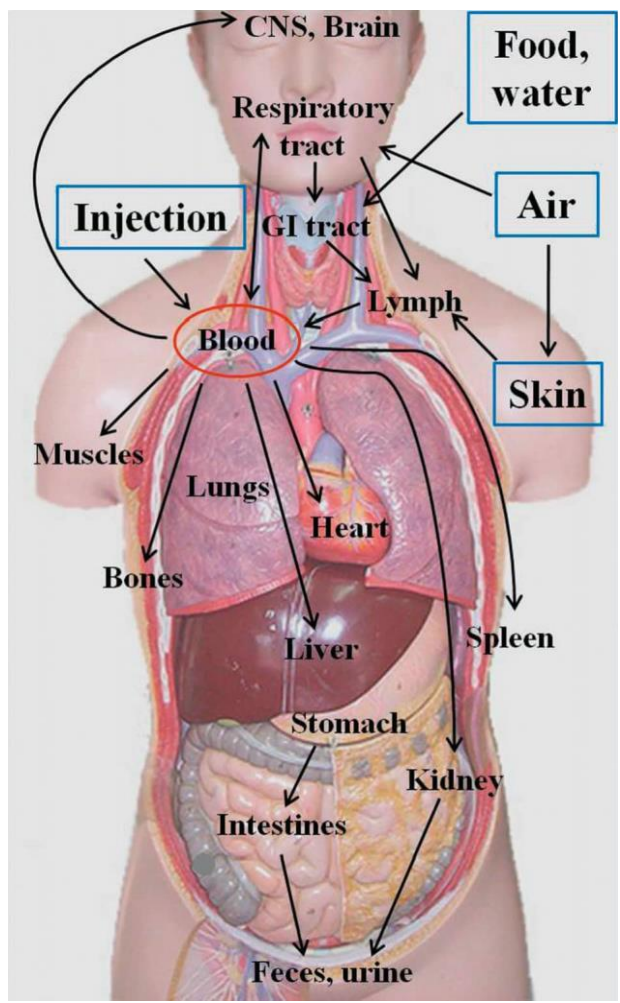 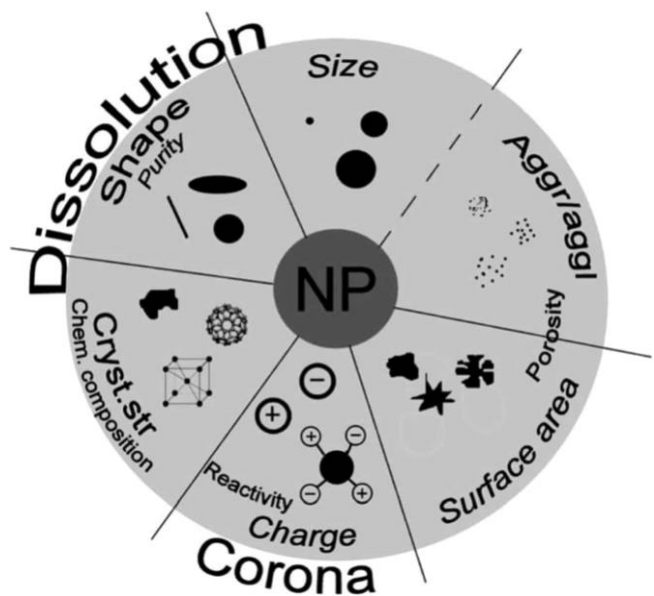

(a)                                        (b)

*Fig 2. (a) Potential routes of exposure, translocation and deposition sites of nanomaterials[22]. (b) Impact of physico-chemical characteristics on nanoparticle-biological system interaction.[23]*

Among different nanomaterials, carbon nanomaterials are revolutionary due to the unique and significant utilization of each of the allotrope. Thus within the scope of the book, the toxicity of carbon nanomaterials will be discussed in this chapter.

## 2. Carbon Nanomaterials

Carbon, placed at group 14 (IV A), can form $sp^3$ and $sp^2$ hybrids and stable multiple pi and sigma bonds which in-turn help it to configure its nano-allotropes like fullerene, carbon nanotube and graphene (Fig 3a and 3b). All of these nano-allotropes possess great promises in finding applications in different potential sectors (Fig 3c)[24].

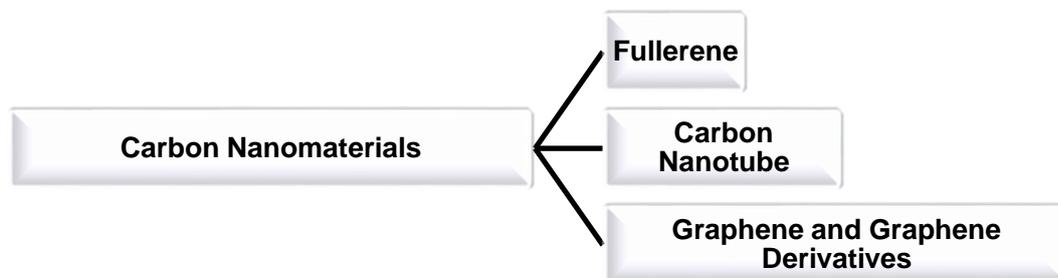

(a)

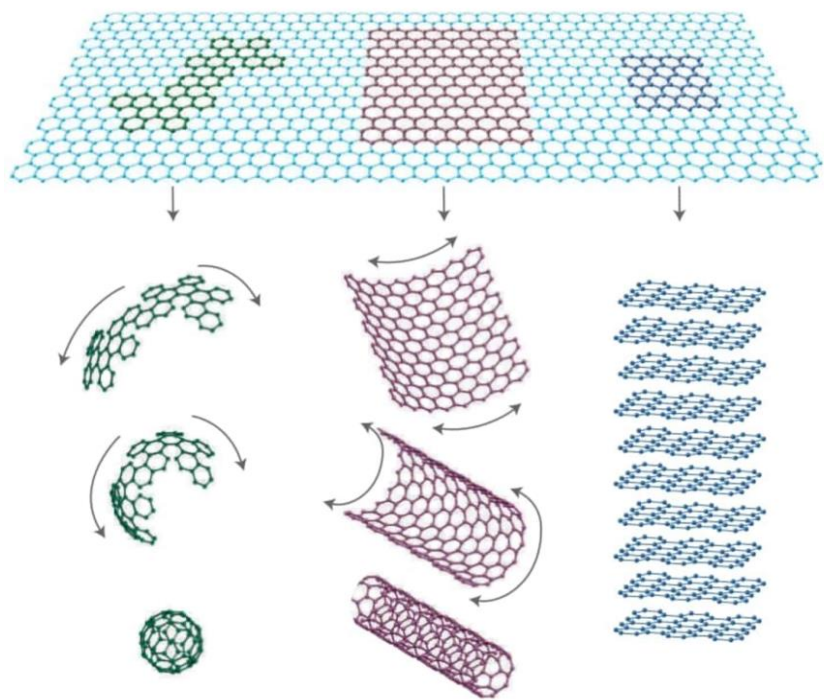

*(b)*

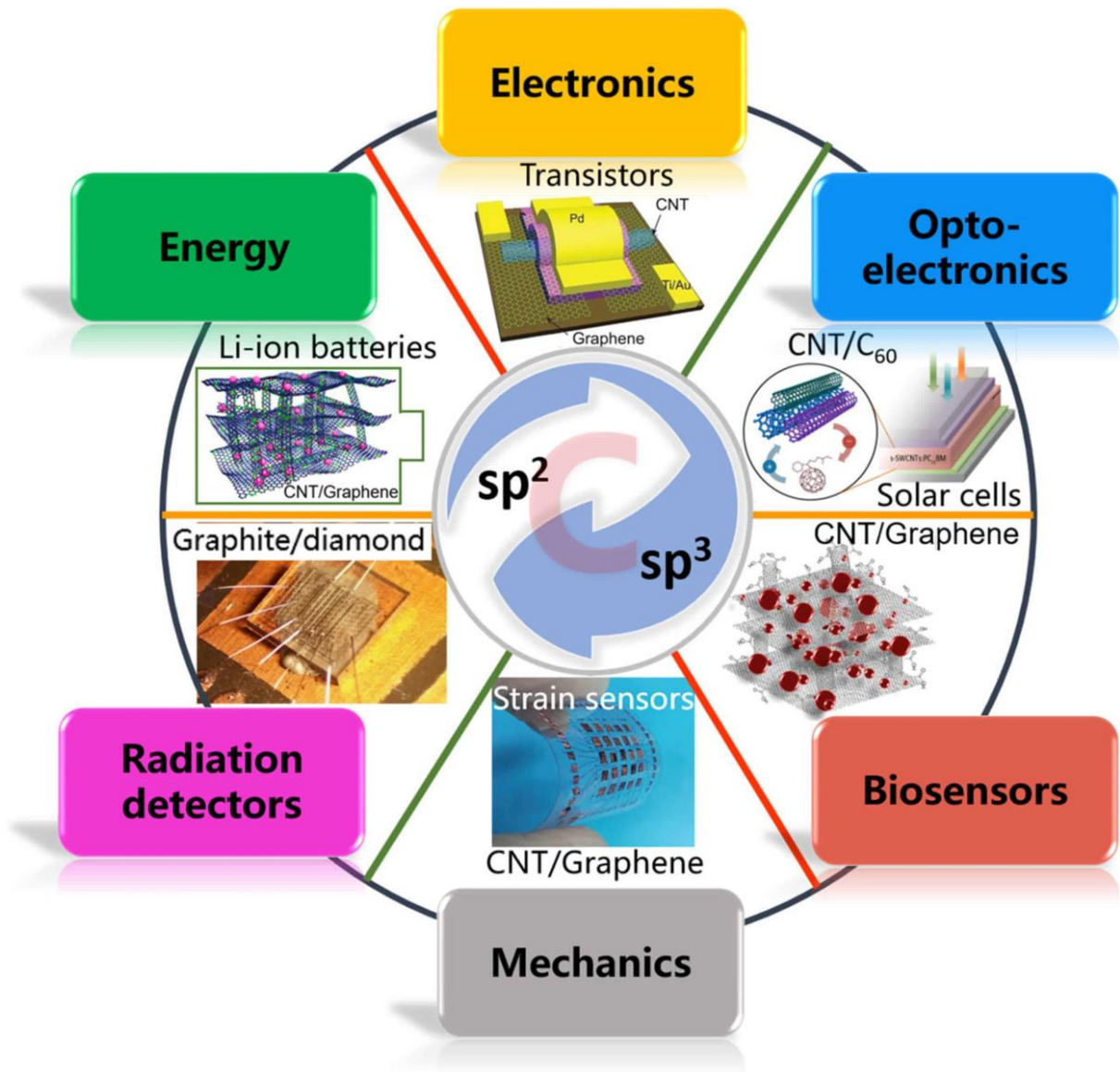

*(c)*

*Fig.3. (a) Different carbon nanomaterials and (b) their structures[25],(c) Different potential application sectors of carbon nanomaterials[26].*

## 2.1 Fullerene

In 1985 Kroto et al.[27] discovered a nearly spherical closed cage molecule having 60 numbers of carbon atoms and named then buckminsterfullerene. Structurally, $C_{60}$ is a type of polygon called truncated icosahedron, comprised of 60 vertices and 32 faces (12 pentagons and 20 hexagons)[28]. The $C_{60}$ molecule possesses high symmetry with numerous transformations each one of which maps the molecule onto itself[29]. A class of water-soluble, oxidized fullerenes, are termed as fullerols or fullerenols. Fullerene can be synthesised by laser vaporization method[30], resistive and inductive heating[31], arc discharge[32], pyrolysis[33] or combustion/ flame synthesis[34]. There are various prospective application areas of fullerene such as fuel cells[35], biology[36] and medicine[37], organic photovoltaics/solar cells[38] etc.

**2.2 Carbon Nanotube**

Radushkevich and Lukyanovich[39] observed a tubular carbon nanostructure in 1952 which was later termed as the carbon nanotube. However, this observation remained unnoticed by the researchers for a few decades. In 1991 Iijima[40] reported the observation of CNTs in the Nature journal, which created global interest and excitement. Structurally, CNT is a rolled-up sheet of carbon atoms that forms the tube structure where the number of sheets may be one (single-wall carbon nanotube or SWCNT) or more (multi-walled carbon nanotube or MWCNT)[41]. The difference in roll-up direction gives rise to different structures of SWCNTs such as zigzag, armchair or chiral[42]. MWCNT can be viewed as either a single sheet of carbon, rolled-in around itself multiple times (Parchment model) or multiple layers of graphitic sheets rolled-in on themselves (Swiss roll). CNTs can be produced primarily via laser ablation[43], arc discharge[44] or chemical vapor deposition[45] methods. There are several possible application areas of CNT such as wearable electronics[46], flat panel display[47], solar cell[48], sensor[49], textile industry[50], Li-ion battery[51], and even in stimuli-responsive material[52].

**2.3 Graphene and Graphene Derivatives**

In 1962, first experimental observation of single layer of graphite was reported by Boehm et al.[53] and in 1994 he named the single layer of graphite as "Graphene"[54]. In 2004, the report of the observation of graphene by Novoselov et al.[55] in Science journal fostered a huge interest in graphene among the global researchers. Structurally, graphene is one atom thick two-dimensional allotrope of carbon, consisting of a single sheet of sp$^2$ hybridised carbon atoms. Graphene can be synthesised using exfoliation methods [56,57], pyrolysis [58], arc discharge[59], epitaxial growth on SiC[60] or chemical vapour deposition[61]. Graphene oxide (GO) is produced by the exfoliation of graphite oxide and can be chemically reduced at a later stage to synthesise graphene. As complete reduction cannot be achieved, the end material is called 'reduced graphene oxide' (rGO). Graphene has a gigantic possibility for applications in different sectors such as energy storage[62], flexible electronics[63], aerospace[64] and biomedical[65] applications.

## 3. Nanotoxicology and Resulting Cytotoxicity or Cellular Toxicity

Nanotoxicology is the branch of biological science which aims to measure the degree to which nanomaterials can cause a hazard to human health (Fig 4). The large surface area-to-volume ratio and tiny size (which gives rise to quantum size effects) of nanomaterials along with environmental factors collectively result in biological changes which widely differs from those caused by their larger counterparts and ultimately cause toxicity. Nanotoxicity can be measured in various ways; however, the primary measurement that needs to be performed is cytotoxicity.

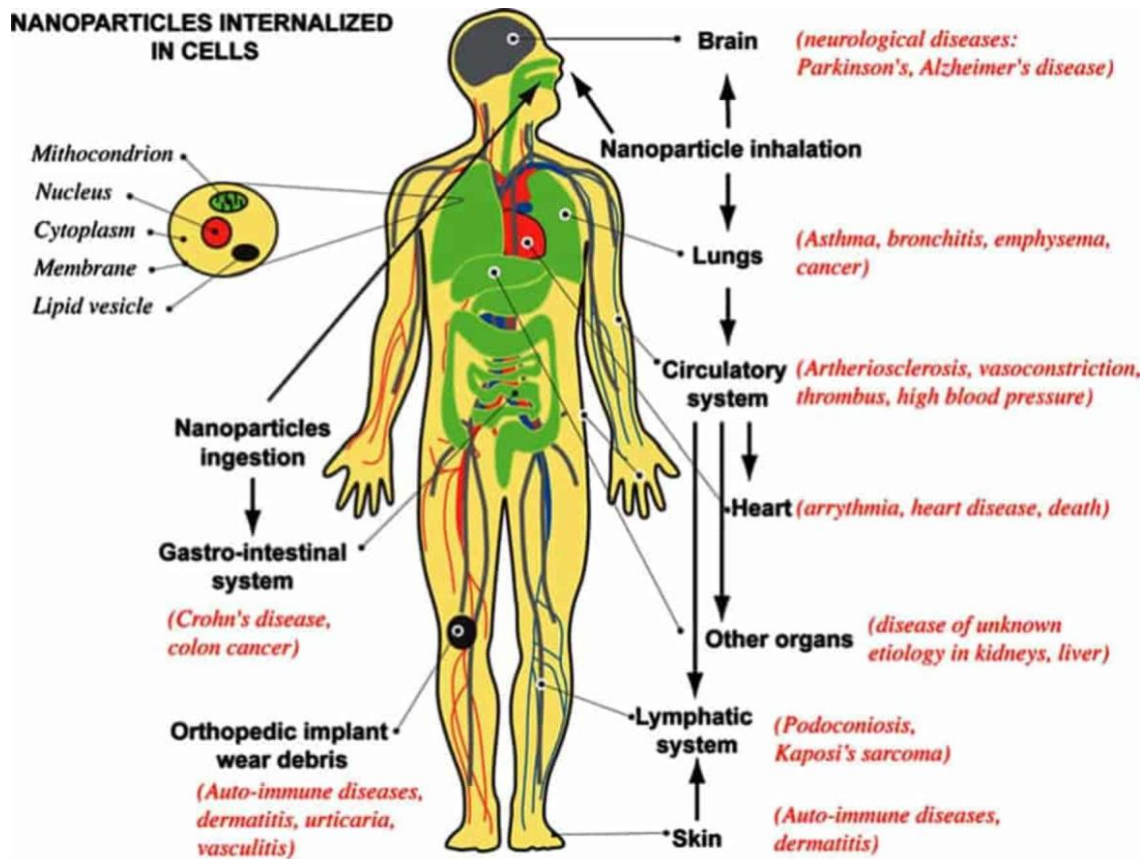

*Fig 4. Pathways of exposure of the human body to nanoparticles, affected organs, and associated diseases[66].*

Cytotoxicity or Cellular toxicity is defined as the potential of a material to cause cell death. For an exact evaluation of cytotoxicity, information about cell origin, cell concentration, drug concentration and treatment tenure require special attention. There are various underlying mechanisms of cytotoxicity, namely, reactive oxygen species[67] (ROS), cell autophagy[68], pyroptosis[69], apoptosis[70] and necrosis[71] (Fig 5).

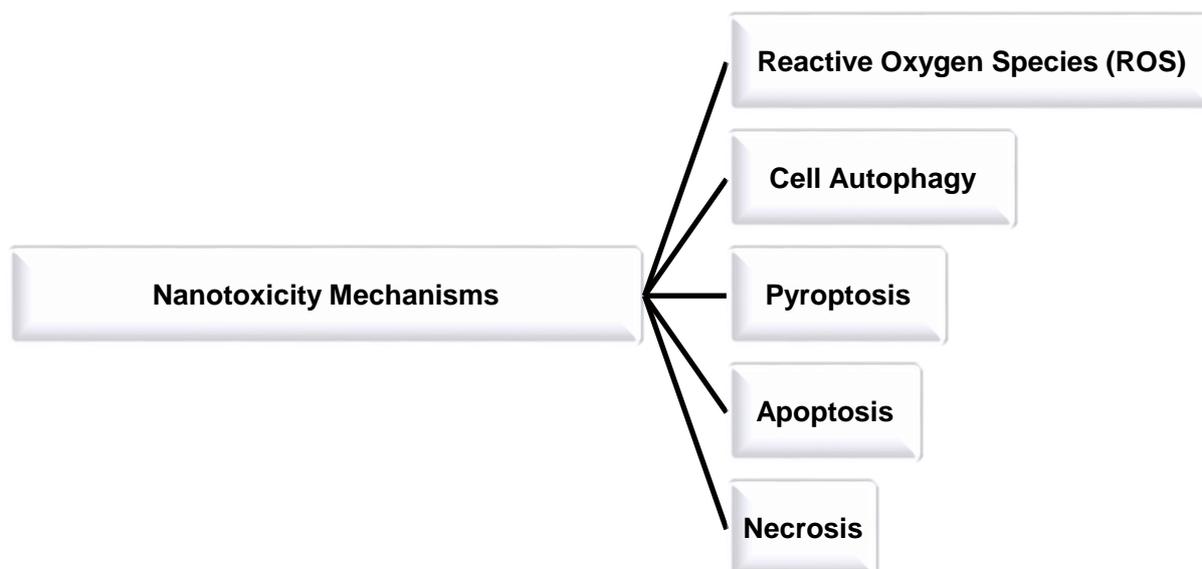

*Fig 5. Various underlying mechanisms of cytotoxicity.*

ROS is a group of oxidative species (Table 1) which are produced intrinsically or extrinsically within the cell. ROS are mostly produced as by-products at the time of mitochondrial electron transport. Additionally, metal-catalyzed oxidation reactions also generate ROS as an essential intermediate. Abundant ROS have the potential to impair biological responses leading to oxidative stress. Excessive toxic levels of oxidative stress can cause electron transfer chain dysfunction along with mitochondrial membrane damage resulting in cell death (Fig 6a).

| Reactive Oxygen Species | Mechanism of Generation |
| --- | --- |
| Superoxide ($O^-_2$) | Reduction of molecular oxygen in the electron transport chain of mitochondria, and other enzymatic routes: monooxygenase, NADPH oxidase, xanthine oxidase |
| Hydrogen peroxide ($H_2O_2$) | Converted from $O^-_2$ by enzyme superoxide dismutase (SOD) |
| Hydroxyl radical ($^•OH$) | Produced in Haber-Weiss reaction from $O^-_2$ and $H_2O_2$ |
| Singlet oxygen ($^1O_2$) | Produced in the reaction of hypochlorous acid (HOCl) and $H_2O_2$ |
| Peroxynitrite ($ONOO^-$) | Produced in the reaction of nitric oxide (NO) |

| | and ($O^-_2$) |
|---|---|

Table 1. Production pathways of Reactive Oxygen Species[72].

Autophagy constitutes the processes of autophagosome synthesis and lysosomal degradation (Fig 6b). Under stressed conditions, reduced lysosomal activity may increase autophagosome synthesis, thus processing of excessive autophagosomes will be reduced due to the rate-limiting lysosomal activity. The formation/accumulation of uncontrolled autophagosomes, which are unfused to lysosomes, leads to autophagy and directly induces cellular toxicity.

The term 'Pyroptosis' originated from the Greek words "pyro" and "ptosis" which mean 'fever/inflammation', and 'to fall' respectively. It is a kind of regulated cell death caused by inflammatory stimuli, transduced via inflammatory caspases and executed by gasdermins. Inflammatory caspases include CASP-1, CASP-4/-5 (human), and CASP-11 (murine) (Fig 6c).

In multi-cellular organisms, the unwanted cells (not needed or threat to the organism) are abolished by a controlled cell suicide procedure known as apoptosis (originated from the Greek word which means "falling off," as leaves from a tree). Proteolytic enzymes (caspases) mediate apoptosis and initiate cell death via cleaving specific proteins in the cytoplasm and nucleus. The members of B-cell lymphoma 2 and Inhibitors of apoptosis protein families regulate the activation process of caspase. (Fig 6d).

Necrosis (originated from the Greek word which means "death") is a type of cell injury which causes premature cell death in living tissue via autolysis (Fig 6e). It is caused by external (to cell or tissue) factors namely trauma, infection, or toxins and leads to the unregulated digestion of cell components.

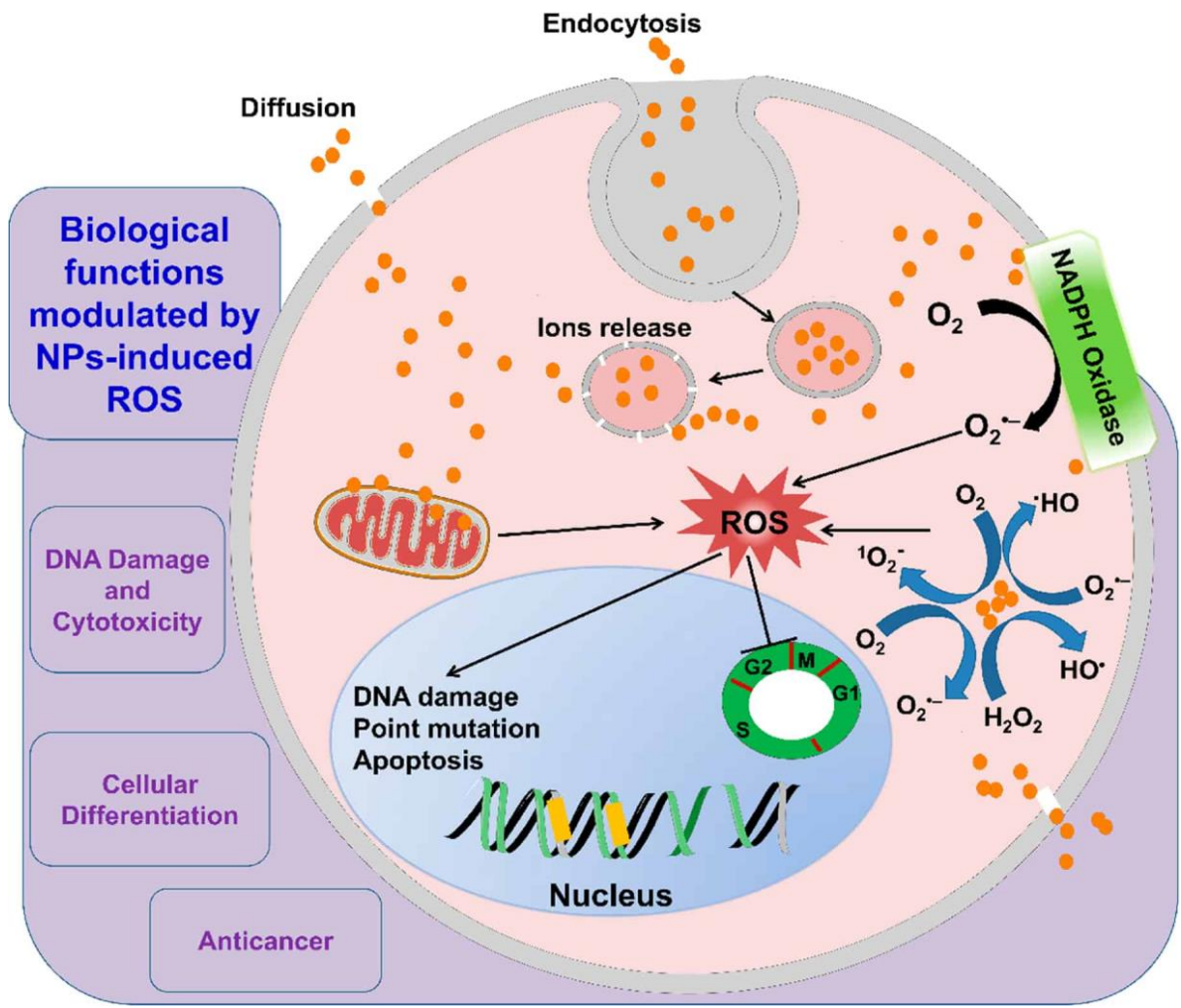

(a)

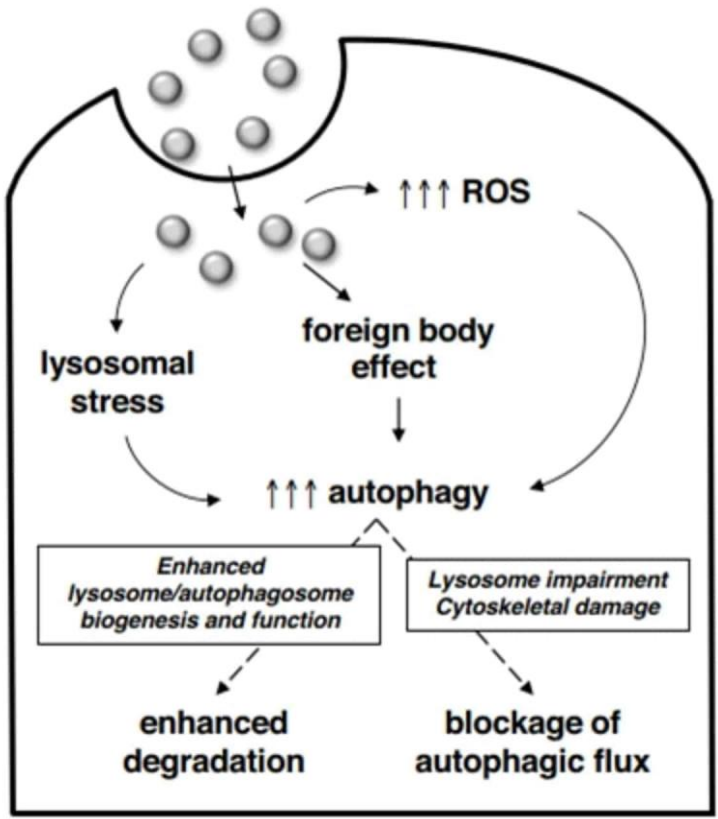

(b)

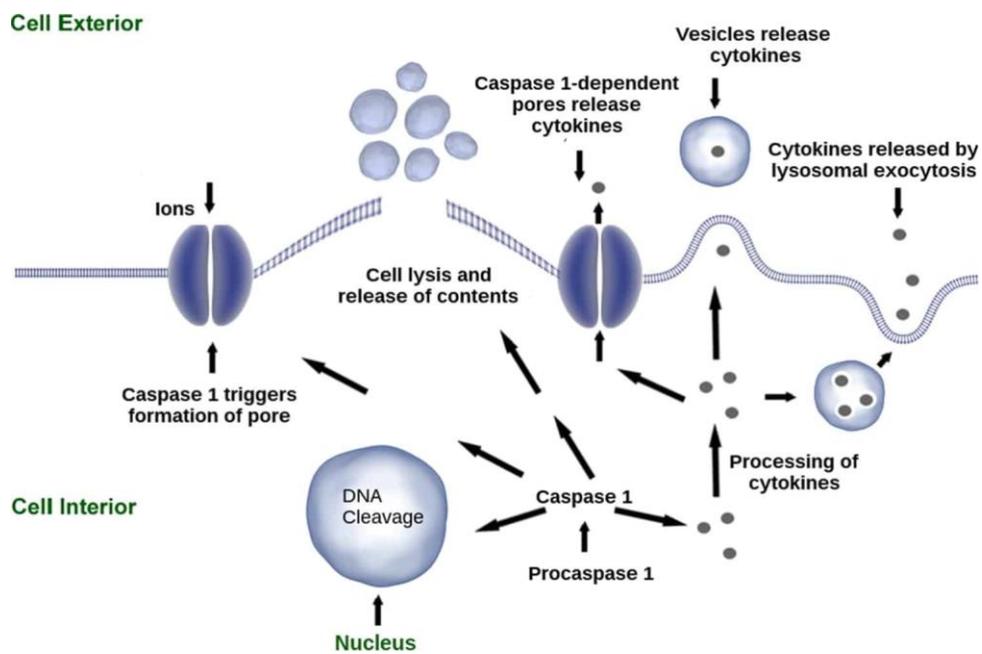

(c)

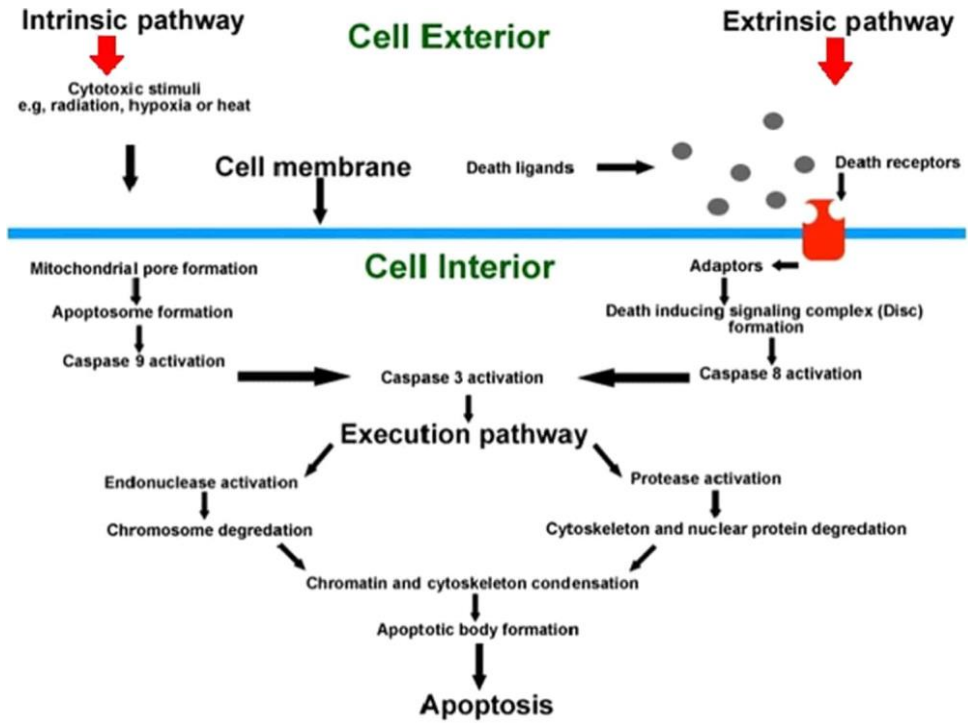

(d)

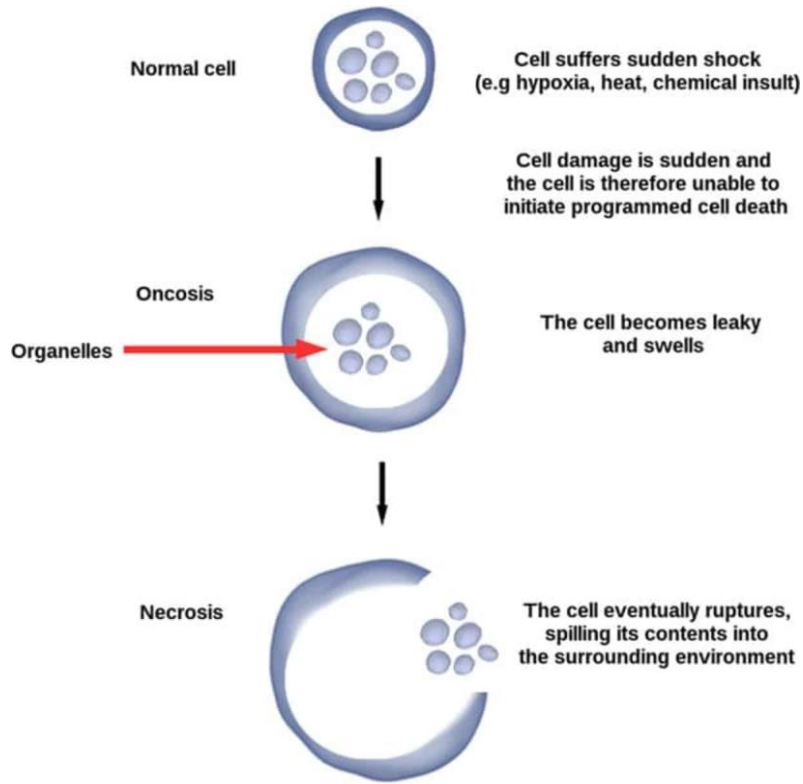

(e)

*Fig. 6. (a) Effect of reactive oxygen species in cells.[73] (b) Mechanisms of nanomaterial-induced autophagy activation[74]. Pathways leading to cell death[75]: (c) pyroptosis, (d) apoptosis, and (e) necrosis.*

The cytotoxicity of carbon-based nanomaterials influenced by physical and chemical properties of carbon nanomaterials[76](Fig 7), surface modification of nanomaterials[77], and metal impurities[78].

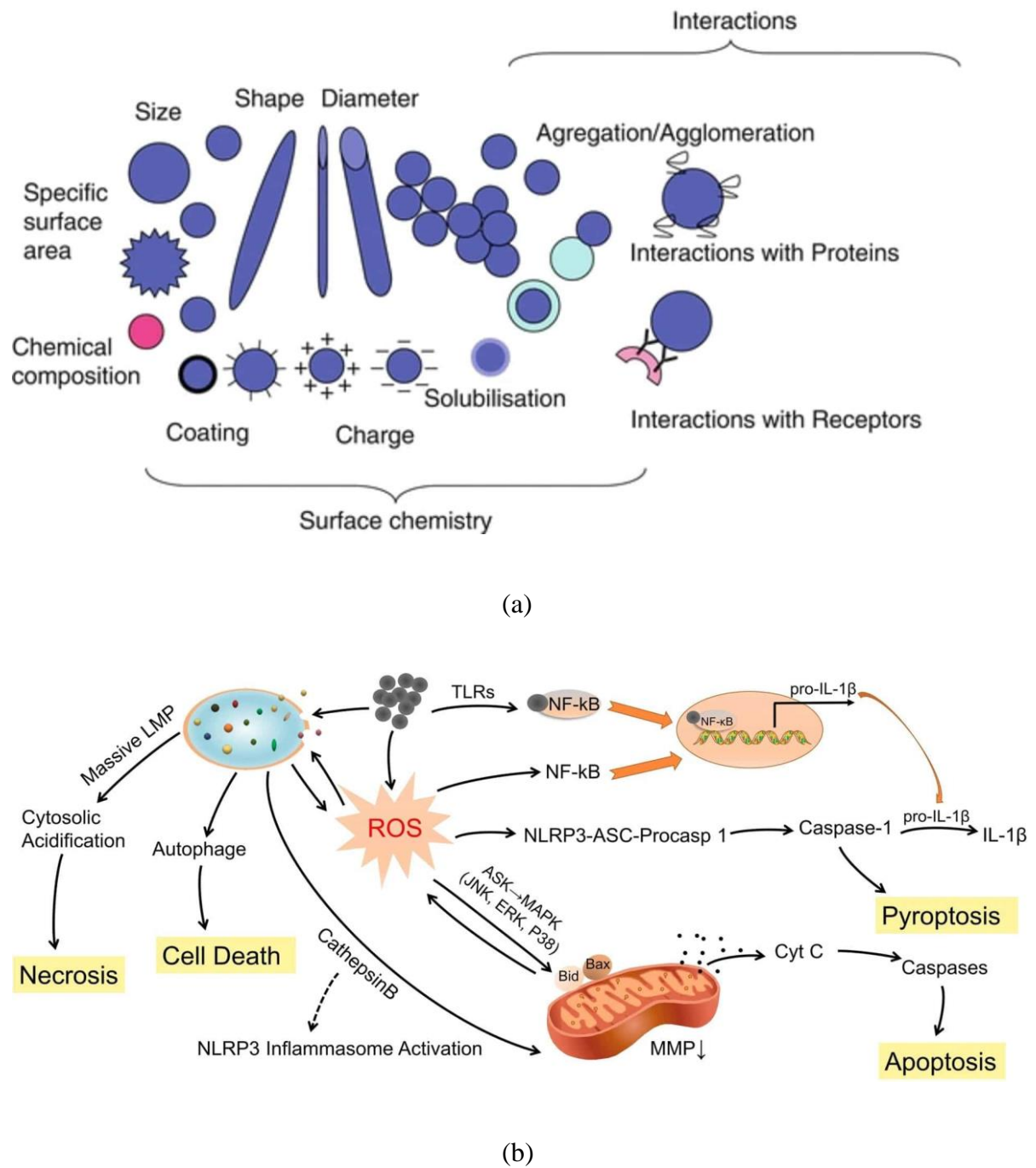

(a)

(b)

*Fig.7. (a) Different physicochemical characteristics of nanomaterials, responsible for their toxicological activity[79]. (b) Cytotoxicity initiation mechanisms of a nanoparticle[80].*

## 4. Assessment of Nano- cytotoxicity

Now cytotoxicity can be assessed by examining the toxicity in different organs of the human body. This section aims to review the cytotoxic effects of carbon nanomaterials in different organs of the human body (Fig 8).

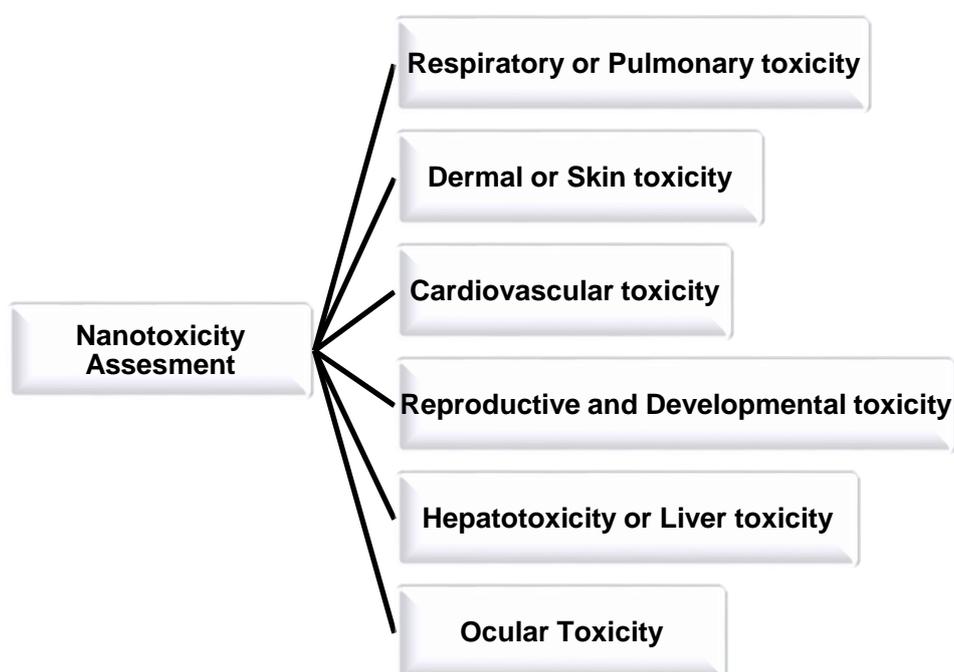

*Fig 8. Nanotoxicity assessment pathways.*

### 4.1 Respiratory or Pulmonary Toxicity

An aggregated form of underivatized $C_{60}$ and a fully derivatized, highly water-soluble derivative of $C_{60}$ was used to study the pulmonary toxicity on rats by Sayes et al.[81]. However, no adverse effects on lung tissue were reported. Two different sized (nano and micro) fullerene aggregates were used by Sayers et al.[82] to study respiratory toxicity in mice and lung inflammatory responses were observed. Morimoto et al.[83] also studied inhalation exposure as well as intratracheal instillation effects of fullerene on rats and their result showed that well-dispersed fullerenes did not cause any neutrophil inflammation.

CNT mediated pulmonary toxicity can occur via inflammation, injury, fibrosis, and pulmonary tumors. Park et al.[84] treated mice with MWCNTs by intratracheal instillation and observed allergic and pro-inflammatory responses owing to B cell (B lymphocytes) activation and production of IgE (Immunoglobulin E). Inoue et al.[85] examined in-vivo and in-vitro effects of pulmonary exposure of MWCNT on mice and showed that MWCNT can intensify allergic airway inflammation with increased humoral immunity and thus has the potential to cause allergic asthma. The effect of short-term inhalation of the rod-like CNT on mice was studied by Rydman et al.[86] and they found that novel natural immunity-mediated allergic-like airway inflammation was induced in mice by the said exposure. Heish et al.[87] studied the potential toxic effects of SWCNT exposure on the respiratory system of mice and found that intratracheal instillation of SWCNTs could result in an airflow obstruction and airway hyperreactivity which could consequently lead to irreversible obstructive airway disease. Xu et al.[88] reported that the intrapulmonary administration of MWCNTs can induce mesothelial proliferation which is closely associated with mesothelioma development. It was reported that myofibroblasts and lung fibrosis can be caused by CNT exposure[89]. Chang et al.[90] reported that epithelial-mesenchymal transition results in SWCNT-induced pulmonary fibrosis. The investigation of Luanpitpong et al.[91] concluded that chronic (6-month) exposure of human lung epithelial cells to SWCNTs can cause lung cancer. Inhalation exposure to MWCNTs can also promote lung adenocarcinoma[92] and lung carcinogenicity[93].

The pulmonary toxicities of different sized graphite nanoplates were explored upon exposure to mice by Roberts et al.[94] which revealed that graphite nanoplates may induce lung inflammation and injury in lavage fluid depending on their sizes. Mao et al.[95] analyzed the toxicity of few-layer graphene in mice after intratracheal instillation which proved that it can cause acute lung injury and pulmonary edema. Ma et al.[96] found that the activation of macrophages and the stimulation of pro-inflammatory responses are governed by the lateral size of graphene oxide. Hu et al.[97] performed theoretical studies to demonstrate that inhaled graphene oxide nanosheets can severely damage pulmonary surfactant film and thereby induce toxicity. Zhang et al.[98] reported that graphene oxide, which is deposited on the lungs, has long retention time and thereby can cause pulmonary edema. Park et al.[99] demonstrated that intratracheally instilled graphene nanoplatelets can remain in lung for long duration and can cause cytoskeletal damage in the lung. Intratracheally instilled functionalised graphene nanoplatelets results in significant acute neutrophilic inflammation as reported by Lee et al.[100]. Nanoscale GO sheets were intratracheally instilled in mice by Li et al.[101] and the study revealed that the incorporation of GO causes acute lung injury and chronic pulmonary fibrosis.

**4.2 Dermal or Skin Toxicity**

Human epidermal keratinocytes were exposed to functionalized fullerene by Rouse et al.[102] which decreased cell viability and initiated a pro-inflammatory response. Saathoff et al.[103] reported that high dose of $C_{60}(OH)_{32}$ (42.5 µg/ml) decreased the viability of human epidermal keratinocytes.

Patlolla et al.[104] applied different doses of purified MWCNT in normal human dermal fibroblast cells to study the cellular response. They found that MWCNTs are very toxic at sufficiently high concentrations. Mice skin was exposed to SWCNT by Murry et al.[105] which resulted in dermal toxicity via enhancement of dermal cell numbers and skin thickening. Zhang et al.[106] studied the effect of the exposure of functionalised SWCNT on human epidermal keratinocytes and conclude that lower concentration of SWCNT can induce mild cytotoxicity, while a higher concentration of SWCNT demonstrated an irritation response. Human epidermal keratinocytes were used as a target for the exposure of unrefined SWCNT and this caused dermal toxicity caused by accelerated oxidative stress as reported by Shvedova et al[107]. Palmer et al.[108] examined the effect of exposure of carboxylated MWCNT on mice and the study revealed that functionalised MWCNT can increase epidermal thickness and gives rise to allergic skin conditions.

As reported by Liao et al.[109], mammalian fibroblasts can be damaged via the exposure to graphene though the exposure environment and mode of interaction with cells are to be considered as crucial parameters,. Pelin et al.[110] investigated differential cytotoxic effects of graphene and GO on skin keratinocytes and found that these materials are cytotoxic and can cause significant cellular damage. In a follow-up study by the same group, they demonstrated that ROS production in human skin keratinocytes could be induced by graphene and GO, which is responsible for their cytotoxic effects.

**4.3 Cardiovascular Toxicity**

To examine the potential vascular toxicity of fullerene in humans, Yamawaki et al.[111] investigated the effects of water-soluble fullerene on vascular endothelial cells and reported that fullerene can cause cytotoxic injury. Thompson et al.[112] found that exposure of fullerene to rats may enhance cardiac ischemia/reperfusion injury via narrowing the coronary artery. Adverse effects of fullerenes on the human umbilical vein endothelial cells were reported by Gelderman et al.[113]. Vesterdal et al.[114] observed that even low doses of fullerene exposure to mice can cause modest vasomotor dysfunction and different degree of atherosclerosis.
Radomski et al.[115] reported that CNT is capable of stimulating platelet aggregation thereby increasing the rate of vascular thrombosis. The cardiac effects of exposure of carbon nanotubes in rat were evaluated by Hosseinpour et al.[116] and their study evidenced that nanotube exposure can induce increment of heart rate. Li et al.[117] demonstrated that intrapharyngeal instillation of SWCNT in mice is capable to cause cardiovascular diseases. Both systolic and diastolic blood pressures were increased and heart rate was decreased on inhalation of MWCNT as reported by Zheng et al.[118].

Singh et al.[119] reported that atomically thin graphene oxide sheets exerted strong aggregatory reaction in platelets depicting its prothrombotic property. Myocardial cells were used to check the cytotoxicity of reduced GO by Contreras-Torres et al.[120] and reported that reduced GO is very toxic to myocardial cells. Bangeppagari et al.[121] exposed zebrafish to graphene oxide and detected several cardiovascular defects due to the exposure. Cardiotoxicity of graphene oxide was evaluated by Arbo et al.[122] via mitochondrial disturbances.

## 4.4 Reproductive and Developmental Toxicity

Tsuchiya et al.[123] exposed mouse embryos to fullerene and after an exposure of 18 hrs, all embryos died. Fullerene can cause oxidative stress in embryonic zebrafish leading to mortality as reported by Usenko et al.[124], however, reduced light during exposure can reduce the mortality rate[125]. Zhu et al.[126] studied the toxicity effects of fullerene aggregates on zebrafish embryos and found that the exposure hindered the development of zebrafish embryo and larva, reducing the survival and hatching rates, and resulted in pericardial edema. Fullerene aggregates were exposed to the Chinese hamster ovary by Han et al.[127] and found that cell mortality increases with the increase in the fullerene concentration and incubation time.

Pietroiusti et al.[128] reported that the application of even a low dose of SWCNT to embryonic stem cell of the mouse can affect embryonic development. Chicken embryos were exposed to SWCNT by Roman et al.[129] found that most of the embryos died before incubation. Fujitani et al.[130] studied the teratogenicity of MWCNT in mice and observed various types of fetuses with skeletal malformation. Cheng et al.[131] assessed the exposure effect of MWCNT on zebrafish embryos and demonstrated that the developmental toxicity is dependent on the length of MWCNT. The effect of functionalized MWCNTs on pregnant mice was studied by Huang et al.[132] and the study revealed that large MWCNT could restrict the foetuses development, and also could induce brain deformity.

Xu et al.[133] injected mouse dams with reduced graphene oxide nanosheets at different doses and time, (pre- or post-fertilization). Reduced graphene oxide nanosheets injected at a late gestational stage is reported to cause abortions and death. Graphene oxide can induce developmental malformation of the cardiac/yolk sac edema, eye, tail flexure and reduction in heart rate when it is introduced during zebrafish embryogenesis, as reported by Chen et al.[134]. Fu et al.[135] reported that graphene oxide can impart many negative effects on the development of mice in the lactation period. Akahavn et al.[136] intravenously injected nanoscale graphene oxide sheets into male mice which caused a reduction in sperm viability and motility.

**4.5 Hepatotoxicity or Liver Toxicity**

Kamat et al.[137] introduced fullerene in rat liver microsomes while exposed to light which resulted in significant lipid peroxidation and other forms of oxidative damage causing liver injury. Cytotoxic effects of hydroxylated fullerenes on isolated rat hepatocytes were studied by Nakagawa et al.[138] and concluded that the exposure can cause cell death via mitochondrial dysfunction. Shipkowski et al.[139] intratracheally instilled fullerene in rats and found that fullerenes are mostly concentrated in the liver and the concentration remains nearly unchanged resulting in detrimental health effects.

Differential cytotoxicity of water-soluble fullerenes on human liver carcinoma cells (HepG2) was investigated by Sayes et al.[140] and found that cytotoxicity of water-soluble fullerene species is governed by surface derivatization. Patlolla et al.[141] demonstrated that the introduction of functionalized SWCNTs in mice causes oxidative stress in the liver followed by morphological alterations of the liver. Awasthi et al.[142] orally administered mice with MWCNTs which resulted in cellular swelling, inflammation and blood coagulation in the liver. Systemic distribution of SWCNT in mice was studied by Principi et al.[143] and the study revealed that the SWCNTs get accumulated in the liver causing liver damage.

Patlolla et al.[144] exposed rats to graphene oxide and found that graphene oxide cause damage to the liver tissue via ROS induction. To study the biological effects of GOs on parenchymal hepatocytes Zhang et al.[145] exposed mouse to GOs and concluded that large GOs cause ROS production and thereby can damage the liver. Mohamed et al.[146] injected graphene oxide nanosheets in male albino mice and observed that graphene oxide has hepatotoxic effects on liver cells via the creation of oxidative stress. The effect of exposure of GO on zebrafish was analyzed by Xiong et al.[147] and they reported that GO could induce liver dysfunction to trigger the oxidative stress along with inflammatory responses to cause liver injury.

**4.6 Ocular Toxicity**

Aoshima et al.[148] observed that highly purified fullerenes can cause corneal epithelial defects and conjunctival redness upon exposure. Cytotoxicity of fullerol in human lens epithelial cells was examined by Roberts et al.[149] and their study revealed that exposure to fullerol along with the presence of sunlight may cause early cataractogenesis. In a similar study performed by Wielgus et al.[150] it was evidenced that, in the presence of sunlight, ocular exposure to fullerol can result in retinal damage.

Ema et al. investigated ocular toxicity on fullerene[151] and carbon nanotubes[152]. The study indicated that both fullerene and carbon nanotube can cause acute irritation of the eyes. Rabbits were exposed to carbon nanotubes to assess their ocular toxicity by Kishore et al.[153]

and upon exposure reversible conjunctival redness and discharge from eye had been reported. Liu's group performed ocular toxicity studies for both SWCNTs[154] and MWCNTs[155] and concluded that CNTs are toxic to human ocular cells. Ocular exposure to SWCNTs and MWCNTs can decrease in the cell survival rate and initiate cell apoptosis via ROS generation.

In order to assess the toxicity of GO exposure to the eye, Wu et al.[156] performed short-term GO exposure on human epithelium cells, which showed significant cytotoxicity with increased intracellular reactive oxygen species. Wu et al.[157] also examined the ocular toxicity of PEG-GO samples having different oxidation levels and found that oxidative stress-induced cytotoxicity of the PEG-GO samples depends on their oxidation levels. An et al.[158] found that short-term repeated GO exposure can cause intraocular inflammation and in turn can initiate cell apoptosis in the cornea thus depicting significant cytotoxicity in rat epithelial cells.

## 5. Conclusions

Nanotechnology is progressing with lightning speed, and so its commercial applications. The toxicity of the resulting exposure has to be assessed with utmost care in saving mankind. Among different nanomaterials, carbon nanomaterials stand far ahead because of their tremendous application potential. Thus many researchers around the globe are diligently trying to rightly assess the toxic nature of carbon nanomaterials. The researchers are using different methodologies and standards for the evaluation of the toxicity of carbon nanomaterials, which possesses difficulty in making a comparative statement based on their work. However, it can be stated that the physical structure and the chemical nature of the nanomaterials are responsible for their nature of toxicity level. Finally, it is the pre-requisite for the industries to properly assess the risk-factors of any carbon nanomaterials in terms of associated toxicity and accordingly develop proper regulations for their safe applications in concerned industries and healthcare sectors.

---

[1] Richard P. Feynman, 'There's Plenty of Room at the Bottom [Data Storage]', *Journal of Microelectromechanical Systems* 1, no. 1 (March 1992): 60–66, https://doi.org/10.1109/84.128057.

[2] Norio Taniguchi, 'On the Basic Concept of 'Nano-Technology', in *Proceedings of the International Conference on Production Engineering*, vol. 2 (International Conference on Production Engineering, Tokyo, Japan: Japan Society of Precision Engineering, 1974), 18–23.

[3] Vibhu Krishnan Viswanathan et al., 'Nanotechnology in Spine Surgery: A Current Update and Critical Review of the Literature', *World Neurosurgery* 123 (1 March 2019): 142–55, https://doi.org/10.1016/j.wneu.2018.11.035.

[4] Yongbing Sun et al., 'Cancer Nanotechnology: Enhancing Tumor Cell Response to Chemotherapy for Hepatocellular Carcinoma Therapy', *Asian Journal of Pharmaceutical Sciences* 14, no. 6 (1 November 2019): 581–94, https://doi.org/10.1016/j.ajps.2019.04.005.

[5] Richard Olawoyin, 'Nanotechnology: The Future of Fire Safety', *Safety Science* 110 (1 December 2018): 214–21, https://doi.org/10.1016/j.ssci.2018.08.016.

[6] Elena Villena de Francisco and Rosa M. García-Estepa, 'Nanotechnology in the Agrofood Industry', *Journal of Food Engineering* 238 (1 December 2018): 1–11, https://doi.org/10.1016/j.jfoodeng.2018.05.024.

[7] Jinu Mathew, Josny Joy, and Soney C. George, 'Potential Applications of Nanotechnology in Transportation: A Review', *Journal of King Saud University - Science* 31, no. 4 (1 October 2019): 586–94, https://doi.org/10.1016/j.jksus.2018.03.015.

[8] Muhammad Usman et al., 'Nanotechnology in Agriculture: Current Status, Challenges and Future Opportunities', *Science of The Total Environment* 721 (15 June 2020): 137778, https://doi.org/10.1016/j.scitotenv.2020.137778.

[9] J. E. Contreras, E. A. Rodriguez, and J. Taha-Tijerina, 'Nanotechnology Applications for Electrical Transformers—A Review', *Electric Power Systems Research* 143 (1 February 2017): 573–84, https://doi.org/10.1016/j.epsr.2016.10.058.

[10] Muili Feyisitan Fakoya and Subhash Nandlal Shah, 'Emergence of Nanotechnology in the Oil and Gas Industry: Emphasis on the Application of Silica Nanoparticles', *Petroleum* 3, no. 4 (1 December 2017): 391–405, https://doi.org/10.1016/j.petlm.2017.03.001.

[11] Ahmed Kadhim Hussein, 'Applications of Nanotechnology in Renewable Energies—A Comprehensive Overview and Understanding', *Renewable and Sustainable Energy Reviews* 42 (1 February 2015): 460–76, https://doi.org/10.1016/j.rser.2014.10.027.

[12] Marina E. Vance et al., 'Nanotechnology in the Real World: Redeveloping the Nanomaterial Consumer Products Inventory', *Beilstein Journal of Nanotechnology* 6, no. 1 (21 August 2015): 1769–80, https://doi.org/10.3762/bjnano.6.181.

[13] Tianmeng Sun et al., 'Engineered Nanoparticles for Drug Delivery in Cancer Therapy', *Angewandte Chemie International Edition* 53, no. 46 (2014): 12320–64, https://doi.org/10.1002/anie.201403036.

[15] J Ferin et al., 'Increased Pulmonary Toxicity of Ultrafine Particles? I. Particle Clearance, Translocation, Morphology', *Journal of Aerosol Science* 21, no. 3 (1 January 1990): 381–84, https://doi.org/10.1016/0021-8502(90)90064-5.

[16] K. Donaldson et al., 'Nanotoxicology', *Occupational and Environmental Medicine* 61, no. 9 (1 September 2004): 727–28, https://doi.org/10.1136/oem.2004.013243.

[17] Jonathan Wood, 'Profiling the Cellular Response to Nanomaterials: Toxicology', *Materials Today* 8, no. 12, Supplement 1 (1 December 2005): 13, https://doi.org/10.1016/S1369-7021(05)71270-0.

[18] H. R. Espanani et al., 'Toxic Effect of Nano-Zinc Oxide', *Bratislava Medical Journal* 116, no. 10 (2015): 616–20, https://doi.org/10.4149/BLL_2015_119.

[19] Fedora Grande and Paola Tucci, 'Titanium Dioxide Nanoparticles: A Risk for Human Health?', Mini-Reviews in Medicinal Chemistry, 31 May 2016, http://www.eurekaselect.com/140530/article.

[20] Leonid M. Kustov et al., 'Silicon Nanoparticles: Characterization and Toxicity Studies', *Environmental Science: Nano* 5, no. 12 (7 December 2018): 2945–51, https://doi.org/10.1039/C8EN00934A.

[21] Catherine Carnovale et al., 'Identifying Trends in Gold Nanoparticle Toxicity and Uptake: Size, Shape, Capping Ligand, and Biological Corona', *ACS Omega* 4, no. 1 (31 January 2019): 242–56, https://doi.org/10.1021/acsomega.8b03227.